\documentstyle[12pt,fleqn,espcrc1]{article}

\advance\voffset by -1.5 cm
\newcommand{\ccbar}{\mbox{$c \overline{c} $}}
\newcommand{\Y}{\mbox{$\Upsilon$}}
\newcommand{\rs}{\mbox{$\sqrt{s}$}}
\newcommand{\pT}{\mbox{$p_T$}}
\newcommand{\xf}{\mbox{$x_F$}}
\newcommand{\rhog}{\mbox{$\rho_g $}}
\newcommand{\jpsi}{\mbox{$J/\psi$}}
\newcommand{\be}{\begin{equation}}
\newcommand{\ee}{\end{equation}}
\begin{document}
\raggedbottom
\pagestyle{empty}
\begin{flushright}
BU-TH-93/2 \\
TIFR/TH/93-35\\
hep-ph/9308230
\end{flushright}
\vspace{1cm}
\begin{center}
{\Large \bf Sizing up the Nuclear Glue in \\
$J/\psi$-production}\\
\vspace{15mm}
{R.V. Gavai$^{1,a)}$ and R.M. Godbole$^{2,b)}$}\\
\vspace{5mm}
{\em 1. Theory Group, Tata Institute of Fundamental Research,
Bombay 400005, India} \\
{\em 2. Dept. of Physics, Univ. of Bombay, Vidyanagari, Bombay
400098, India}
\\
\end{center}

\vspace{2.cm}
\begin{abstract}

Nuclear gluon densities are of great importance to the
physics of relativistic heavy ion collisions, in particular, in assessing
the origin of $J/\psi$-suppression. We describe our attempts to distinguish
various models of the gluonic EMC-effect, using the existing
$J/\psi$-production  data in proton-nucleus collisions.
We find that no model is capable of explaining {\em all} the features
of the high precision E772 data although the overall trend
suggests this to be more a matter of fine-tuning the models than
the presence of new physical effects.

\end{abstract}
\vspace{1.5cm}
\begin{center}
Talk presented at \\
`Quark Matter '93', Borl\"ange, Sweden \\
June 20-24, 1993\\
\end{center}

\vspace{3cm}
\noindent
$^{a)} $ gavai@theory.tifr.res.in\\
$^{b)} $ rohini@theory.tifr.res.in\\

\vfill

\section{INTRODUCTION}

Theoretical investigations of several aspects of the heavy-ion collisions
at high energies depend crucially on the precise knowledge of the gluon
distributions of the colliding nuclei. A well-known prime example is the
\jpsi-suppression signal\cite{MatSat,KarPet} of quark-gluon plasma (QGP).
It has been shown\cite{GavGup,GaGuSr} that this signal and its transverse
momentum dependence can be mimicked by the nuclear structure function
effects.  The
nuclear glue can thus be a potentially serious source of background which
has to be subtracted off to see QGP.  Another important reason to size up
the nuclear glue is the EMC-effect observed in the ratio of quark structure
functions, $\rho = F^A_2(x)/A F_2(x)$.  There are a variety of diverse
theoretical ideas which ``explain'' the data on $\rho$.  Since most of
these result in a prediction for the corresponding gluon ratio, \rhog{},
any attempt to obtain experimental information on \rhog{} is likely to
assist in distinguishing between the various models and in pinning down the
likely cause of this effect.

Extraction of gluon densities from the data is known to be a difficult
exercise although there are various known methods such as the direct
photon production or the \jpsi-production.  The problem is that in most
cases the quark distributions also contribute and one has to invent
clever tricks or resort to some approximations to separate the gluon
densities.  Recently, Drees and Kim advocated\cite{DreKim} the study
of associated production of \jpsi{} and a photon at large \pT{} as a clean
probe of the gluon density in $pp$ and $ep$ collisions. We extended\cite
{GaGoSr} their considerations to $pA$ and $AA$ collisions for the fixed
target experiments at FNAL and the collider experiments at RHIC.
We found that the rates
for these processes at both FNAL and RHIC are substantial and a large
range of $x$ is covered by combining them, thus enabling a cleaner
and better determination of \rhog{}. For more details, including a technique
to improve the statistics, we refer the reader to Ref.\cite{GaGoSr}.

\section{$J/\psi$-PRODUCTION}

While the associated production of \jpsi{} and photon at large \pT{}
could serve as a good probe of \rhog{} in future, one naturally
asks the question whether the existing data can in some way be
used for this purpose.  In particular, the recent high statistics
\jpsi{} and \Y-production data from the E772 experiment\cite{E772}, combined
with the older data \cite{OldDat} have been exploited by Gupta and
Satz\cite{GupSat} to obtain \rhog{}.  We argue that their assumptions
are not valid and one therefore needs to take a less ambitious approach
of

checking whether the data can be explained by the existing models for
\rhog{}.  Before we do so, let us recall briefly the theoretical formalism.

\subsection{Theoretical Models}

The basic perturbative QCD-subprocesses which contribute to the production
of \jpsi{} are
\be
g g \rightarrow \ccbar ;~ q \bar q \rightarrow \ccbar {\rm ~~and~~}
g g \rightarrow \ccbar g ;~ q \bar q \rightarrow \ccbar g ;~
q g \rightarrow \ccbar q ;~ \bar q g \rightarrow \ccbar \bar q
\ee
The first two of these contribute at $O(\alpha_s^2)$ while the rest do so
at $O(\alpha_s^3)$.  Furthermore, since the \ccbar{} pair will hadronize
to \jpsi{}, it is also clear that the former will result in \jpsi{} at low
($\sim 0$) \pT{} while the latter will yield large \pT{} for \jpsi{}.
Two popular\cite{BaiRue}
models for hadronization of the \ccbar{} pair to \jpsi{} are 1)
semilocal duality and 2) colour singlet model.  The former simply consists
of computing the \ccbar{} pair cross section with the appropriate
perturbative QCD cross sections and the chosen quark and gluon densities
and multiplying them by a constant if the pair mass lies in the range
between $M_{J/\psi}$ and $M_D$.  Hadronization in the latter model, on
the other hand, is governed by the quarkonium
wavefunction (or its derivative) at the origin.  An intersting
feature to note is that the charmonium is produced at the basic QCD vertex
in this model and the kinematics is very different.  Thus, e.g., the
\jpsi{} at large \pT{}  is produced in a basic $ 2 \rightarrow 2$ process in
contrast to the semilocal duality model above.

\subsection{Results}

In order to extract \rhog{}, Ref. \cite{GupSat} assumed that the data
can be described by $O(\alpha_s^2)$ partonic cross section and by further
assuming that gluons dominate.  Since the \pT-range of E772 goes up to
2.25 GeV, the first assumption implies rather high values for
primordial
\pT{}:
$\langle \pT \rangle \sim 1-2 $ GeV.  Furthermore, as Fig. \ref{fig:gludom}
shows, the assumption of gluon dominance is both \xf-dependent and
structure function parametrization-dependent.
What is plotted in the figure is the ratio of the
product of quark structure functions to that of gluon structure functions
which act respectively as the weights of the quark-antiquark and gluon-gluon
terms in $pp$ collisions at $\tau=M_{J/\psi}/\rs = 0.0775$.
The various curves are for different popular
parametrizations of the structure functions. The corresponding curves for
\Y{} cast even more doubt on the validity of gluon dominance.

We therefore used the $O(\alpha_s^3)$ partonic subprocesses to compute the
\pT-distribution of \jpsi{} and \Y{} for the E772 acceptance for both
the semilocal duality model and the colour singlet model.  The details
of our calculations can be found in Ref. \cite{GavGod}.
Here we wish to make only some general comments. Since the E772 data is
unfortunately not sufficiently differential, we had to extend our
integration range over larger \xf{} where our approach may be inadequate.
Fortunately, the contribution from the large \xf{} region is small and
thus will not affect our conclusions.  Since the \xf-spectrum of
E772 also includes low \pT{} data, we are unable to compute it at all in
our approach.  Figs. \ref{fig:carb} and \ref{fig:tung} exhibit our results
along with
the E772 data for their lightest and heaviest target for three
different models of \rhog{}.  Again, more details about these models,
including original references can be found in Ref. \cite{GavGod}. What
is remarkable is that for Tungsten, none of the model agrees with the
data, although for carbon, the so-called Gas model seems to perform
well.  For the other two targets of E772, situation is akin to that
in Fig. \ref{fig:carb}.

Performing the same computation for the colour singlet model is
computationally a lot easier since the final state is simpler. In order
to highlight the independence of the above results on the hadronization
models in a typical example, we display in Fig. \ref{fig:tungs}
the ratio of the results for the two models discussed above as a function
of \pT{} for the Tungsten nucleus.
The agreement is really remarkable, especially when one takes
into account the differences in kinematics as well.  We conclude from here
that the data are really a reflection of the structure function effects
alone, although the models of \rhog{} considered here need to be fine-tuned.

\begin{figure}
\caption{$\alpha(\pT)$ vs. \pT : \Y{} \label{fig:ups}}
\vspace{-1.8cm}
\end{figure}
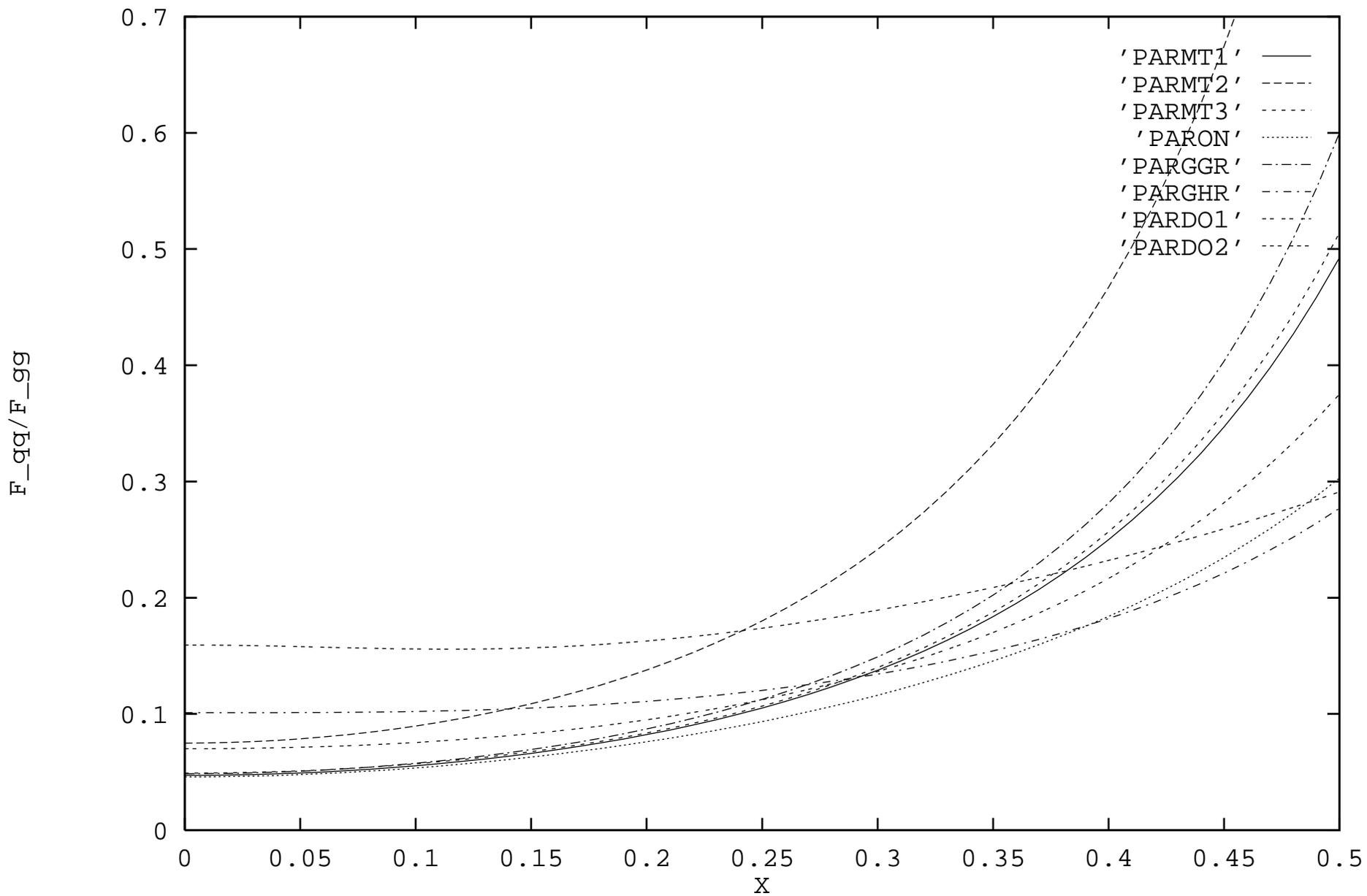
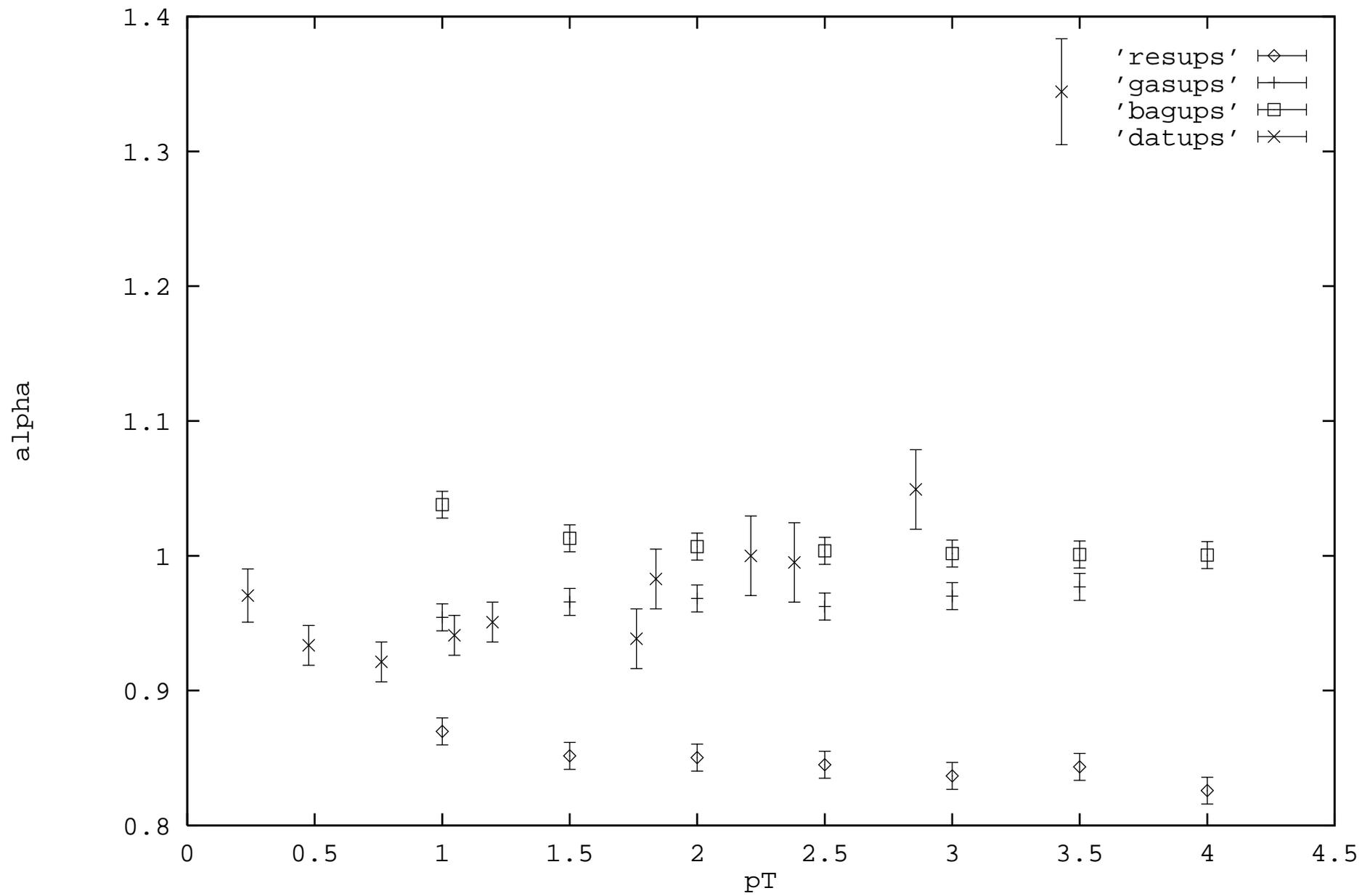

Essentially the same conclusion seems to emerge from our analysis of the
E772 data on \Y{}-production.  Since only $\alpha(\pT)$ as a function
of \pT{} is available in this case, where $\alpha(\pT)$ is obtained
by fitting the cross section at each value of \pT{} for all nuclei to
the form $A^\alpha$, we followed the same E772 procedure and obtained
$\alpha(\pT)$ in the range 1-4 GeV.  As seen in Fig. \ref{fig:ups},
a very good agreement with the E772 data is obtained
for both the Gas model and a 6-quark cluster model for \pT{} upto 3 GeV.
Surprisingly, the computations remain much flatter at even higher
\pT{} values whereas the data show a dramatic increase.  Since the
large \pT{}-domain is theoretically better suited for the description
employed above, this discrepancy is even more striking and needs to be
understood more carefully.

\newpage


\begin{figure}
\centerline{Figure Captions}
\vspace{0.5cm}
\caption{Test of gluon dominance as a functionof$x_F$.\label{fig:gludom}}
\vspace{-1.8cm}
\end{figure}
\vspace{-1cm}
\begin{figure}
\caption{R(\pT) vs. \pT :Carbon \label{fig:carb}}
\vspace{-1.5cm}
\end{figure}
\vspace{-1cm}
\begin{figure}
\caption{R(\pT) vs. \pT :Tungsten \label{fig:tung}}
\vspace{-1.8cm}
\end{figure}
\vspace{-1cm}
\begin{figure}
\caption{ Ratio of models: Tungsten \label{fig:tungs}}
\vspace{-1.8cm}
\end{figure}
\vspace{-1cm}
\end{document}